\documentclass[useAMS,usenatbib,intlimits,centertags]{mn2e}
\usepackage{graphicx}
\usepackage{citesort}
\usepackage{amsmath}
\usepackage{booktabs}

\newcommand{\rdr}{r_1/r_2} 

\begin{document}

 \title{Equilibrium Configurations of Homogeneous Fluids in General Relativity}

 \author[Ansorg, Fischer, Kleinw\"achter, Meinel, Petroff \& Sch\"obel]
 {M.\ Ansorg\thanks{current address: Max-Planck-Institut f\"ur Gra\-vi\-ta\-ti\-ons\-phy\-sik, Albert-Einstein-Institut, Am M\"uhlenberg 1, 14476 Golm, Germany},
 \addtocounter{footnote}{5}
 T.\ Fischer,
 A.\ Kleinw\"{a}chter,
 R.\ Meinel\thanks{Meinel@tpi.uni-jena.de},
 D.\ Petroff and
 K.\ Sch\"{o}bel\\
 Theoretisch-Physikalisches Institut, University of Jena,
                Max-Wien-Platz 1, 07743 Jena, Germany}
 \date{\today}

 \pagerange{\pageref{firstpage}--\pageref{lastpage}} \pubyear{2004}

 \maketitle

 \label{firstpage}
 
 \begin{abstract}
  By means of a highly accurate, multi-domain, pseudo-spectral method, we
investigate the solution space of uniformly rotating, homogeneous and
axisymmetric relativistic fluid bodies. It turns out that this space can be
divided up into classes of solutions. In this paper, we present two new classes
including relativistic core-ring and two-ring solutions. Combining our
knowledge of the first four classes with post-Newtonian results and
the Newtonian portion of the first ten classes, we present the qualitative
behaviour of the entire relativistic solution space. The Newtonian disc limit
can only be reached by going through infinitely many of the aforementioned
classes. Only once this limiting process has been consummated, can one proceed
again into the relativistic regime and arrive at the analytically known
relativistic disc of dust.
 \end{abstract}



 \section{Introduction}

Self-gravitating bodies of constant density have always played a central role
in the physics of gravitation. Contributions that have been most significant to
the aspects of this subject that will be relevant to this paper can be divided
into (1) work done within Newton's theory of gravitation: \cite{Maclaurin},
\cite{Poincare}, \cite{Dyson92,Dyson93}, \cite{Lichtenstein}, \cite{Wong74} and
\cite{Eriguchi81} and (2) work done within Einstein's theory of gravitation:
\cite{Schwarzschild}, \cite{Chand67}, \cite{Bardeen71}, \cite{BI76} and
\cite{GonGour02}.  Despite this great investment of effort, it has not yet been
possible to explore the whole spectrum of Newtonian, let alone relativistic
solutions, even if one restricts oneself to axial symmetry and stationarity.
Using sophisticated computer programs, we believe that a great step can be
taken toward painting the complete relativistic picture of uniformly rotating,
homogeneous and axisymmetric bodies. However, because of the many intricate
details in this picture, particularly in the vicinity of limiting
configurations, it is necessary to use a computer program robust enough to be
able to render such limiting configurations and accurate enough to be able to
distinguish between neighbouring solutions of Einstein's equations. For our
investigations of homogeneous fluids, we used a code based on multi-domain
pseudo-spectral methods \citep{AKM1,AKM3}, which satisfies these requirements.
For a comparison with other codes, see \cite{Stergioulas}.

It is our intention in this paper to present the relativistic picture in its
entirety by describing those parts of it that we have studied explicitly and
augmenting them with conjectures as to the rest. With this in mind, we focus
our attention neither on the numerical methods nor primarily on properties of
individual configurations (as in \citealt{AKM4,SA}), but instead emphasize the
interrelations between various configurations and portray the solution space as
a whole. As an interesting example of the newly explored configurations, we
present the shape and various physical parameters of a core-ring solution.

The picture that emerges for relativistic homogeneous fluids contains familiar
demarcations, which we can use to orient ourselves. It contains, for example,
three analytically known solutions: the (inner and outer) Schwarzschild
solution \citep{Schwarzschild}, the relativistic disc of dust solution
\citep{BW71,NM95,NM03} as well as the Maclaurin solution \citep{Maclaurin} as
one of its Newtonian limits. This last solution will be a particularly useful
point of departure for describing the corresponding relativistic picture. It
represents homogeneous, rotating fluid spheroids and, but for a scaling factor,
depends for given mass density on only one parameter, thereby allowing one to
refer to the ``Maclaurin sequence''. After introducing some basic equations in
Sec.~\ref{Equations} we thus turn our attention to the Maclaurin solution and
its post-Newtonian extension in Sec.~\ref{Maclaurin}.  It turns out that not
all relativistic configurations are connected to each other in a continuous
way, and it will be useful to introduce the notion of a class of solutions. A
given class will be defined to include all configurations of strictly positive
mass that are connected continuously to each other. In Sec.~\ref{Classes} we
review the characteristics of the ``first'' three classes in detail and provide
an overview of the remaining classes.  We close with a discussion of the
limitations of numerical methods and some thoughts on the completeness of the
relativistic picture painted in this paper.

 \section{Einstein's Field Equations and their Numerical
          Solution}\label{Equations}
  Using Lewis-Papapetrou coordinates, the line element for a stationary, axially
symmetric spacetime corresponding to a rigidly rotating perfect 
fluid configuration can be written in the form
\[ds^2=e^{2\alpha}(d\varrho^2+d\zeta^2)+W^2e^{-2\nu}(d\varphi-\omega\,dt)^2-
       e^{2\nu}\,dt^2,\]
where the metric functions $\alpha$, $W$, $\nu$ and $\omega$ are functions of
$\varrho$ and $\zeta$ alone. The boundary of a fluid configuration rotating
with the angular velocity $\Omega$ obeys the equation
\[ e^{2\nu}-W^2(\omega-\Omega)^2e^{-2\nu}=\text{const.}=\left(1+Z_0
\right)^{-2},\]
where $Z_0$ is the relative redshift observed at infinity of a photon emitted
with zero angular momentum from the configuration's surface. Together with
regularity requirements along the axis of rotation, $\varrho=0$, and the 
asymptotic
behaviour, the boundary equation including transition conditions and the field
equations themselves form a complete set of equations to be solved (see e.g.\
\citealt{BI76}). In this paper we choose natural units in which the gravitational
constant $G$ and the speed of light $c$ are both equal to one.

This set of equations was solved numerically to very high accuracy using a
multi-domain pseudo-spectral method. In the algorithm used, the domains are
chosen such that the unknown boundary $\zeta=\zeta_{\text s}(\varrho)$ of the
configuration always coincides with the boundary between two domains, thus
avoiding Gibbs' phenomena. The boundary enters into the equations through
coordinate transformations, which map the physical domains onto a square. The
metric functions and the boundary are represented by a truncated Chebyshev
expansion so that the above mentioned set of equations can be reduced to a set
of (non-linear) algebraic equations. Typically, we used between 10 and 30 Chebyshev coefficients per dimension, depending on the properties of the function being represented. The parameters we can set to specify a fluid configuration include a mass-shedding parameter ($\beta$ of Sec.~\ref{ClassI}) and central pressure, which can be set strictly to infinity (through a rescaling), thereby allowing us to study even the boundary configurations to be discussed in this paper with great accuracy. Further details can be found in
\cite{AKM3}.

 \section{The Maclaurin Sequence and its Bifurcation Points}\label{Maclaurin}
  Maclaurin showed that for a spheroid (also called an ellipsoid of rotation) of
constant density there is a balance of gravitational, centrifugal and pressure
forces, leading to equilibrium in Newtonian theory. For a given mass density,
this spheroid can be described entirely by specifying the ratio of polar to
equatorial radius, 
\begin{equation}\label{eq:A} A:=\rdr,\end{equation}
and the focal distance of the ellipse used to generate
this ellipsoid of revolution. The focal distance merely provides the scaling of
the
spheroid and we can use $A \in [0,1]$ to parametrize the whole solution.

\cite{Poincare}, \cite{Bardeen71} and \cite{HuE} all studied points of instability along the Maclaurin sequence and
the latter three authors were able to show that these correspond to bifurcation
points. The bifurcation points that are of particular interest to us here are
those marking the onset of the various modes of axially symmetric, secular
instability. Values for $A$ at the first few of these can be found in
Table~\ref{bifurcation} and will be denoted by $A_n$, ($n=1,2,3,\ldots$).
\begin{table}
 \caption{Values for the ratio of polar to equatorial radius of the Maclaurin
	  spheroid at the first six bifurcation points associated with the onset
	  of axially symmetric, secular instability. \label{bifurcation}}
 \begin{center}
 \begin{tabular}{c} \toprule
  Bifurcation points \\ \midrule
  $A_1= 0.17126186\ldots$ \\
  $A_2= 0.11160323\ldots$ \\
  $A_3= 0.08274993\ldots$ \\
  $A_4= 0.06574427\ldots$ \\
  $A_5= 0.05453402\ldots$ \\
  $A_6= 0.04658868\ldots$ \\ \bottomrule
 \end{tabular}
 \end{center}
\end{table}
There are countably infinitely many such bifurcation points with the accumulation
point $A=0$.
\cite{Chand67} calculated the post-Newtonian corrections to the
Maclaurin spheroids and found that they become singular at the bifurcation point
$A_1$. \cite{Bardeen71} conjectured that the
$n^{\text{th}}$ post-Newtonian approximation to the Maclaurin spheroids becomes
singular at each of the first $n$ such bifurcation points --  a conjecture
which was then proved in \cite{P03}.

Assuming the convergence of the post-Newtonian series, an assumption based on
strong evidence especially in the disc limit (see \citealt{PM}), the singularities
then represent impermeable barriers in an appropriate parameter space.
These barriers allow us to divide our solution space into ``classes'' of
configurations in the following way: We first divide up the Maclaurin sequence
into an infinite number of segments with the above mentioned bifurcation
points, $A_n$, separating one segment from the next. These segments can
be numbered (or named) starting, for example, from $A=1$. We can then
identify uniquely any configuration of strictly positive mass with a given
segement by requiring that the segment be approachable in a continuous way.
All those configurations that have been thus associated with segment $n$ make up
class $n$.

As we shall see in the following sections, numerical results lend strong
support to the idea that such impermeable barriers as were mentioned above
exist and that the definition of a class is thereby justified. In particular,
we shall see that it is possible to place ourselves at one of the end-points of
a Maclaurin segment and move off the Maclaurin sequence along other well-defined
boundaries until we
land again on the Maclaurin segment at the second end-point. Neighbouring
classes
have only this one point as a common boundary.

 \section{The Relativistic Classes}\label{Classes}
  \subsection{Class I: The Generalized Schwarzschild Class}\label{ClassI}
The first class of configurations that will be discussed in this paper was
explored extensively in \cite{SA}. Here we shall give a brief overview of it,
paying particular attention to its defining boundaries.

For a given constant mass density, $\mu$, a stationary, axisymmetric, rigidly
rotating
relativistic fluid depends on two parameters. A limiting configuration, i.e.\
one that possesses a physical property representing a boundary, is specified
by fixing a corresponding parameter, meaning that it depends on only one further
parameter.
It is therefore possible to refer to a boundary sequence and identify the
end-points of this sequence uniquely with respect to the second parameter.

\begin{figure*}
 \includegraphics[width=\textwidth]{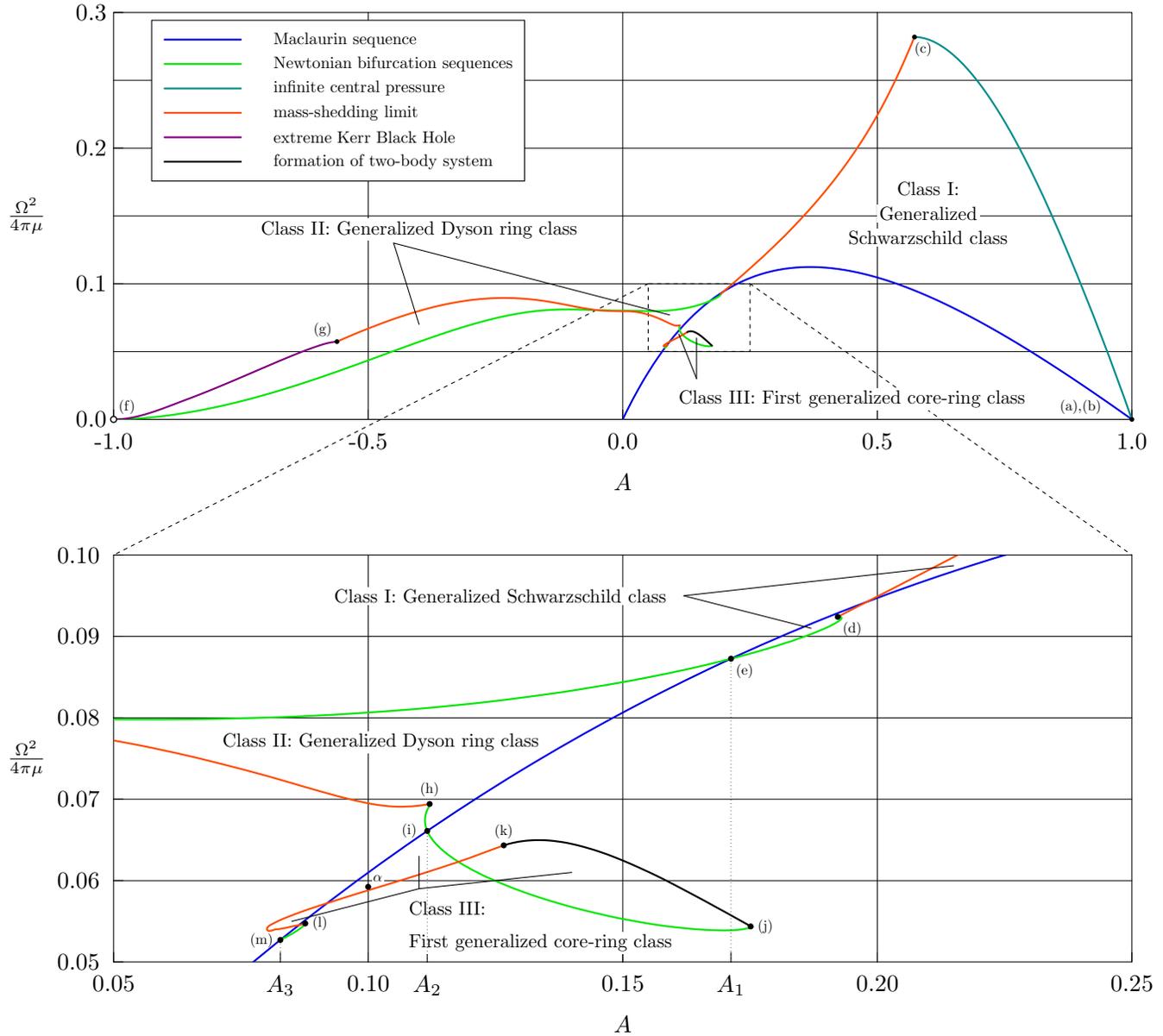}
 \caption{\label{Vollbild}The square of the normalized angular velocity
 versus the ratio of radii for the boundary curves of the
 first three classes.}
\end{figure*}

Clearly the static boundary is marked by the specification $A=1$. 
(Note that we also use the radius ratio as defined in Eq.~\eqref{eq:A} in the
relativistic context, where $r_1$ and $r_2$ are now coordinate radii.)
Static
configurations are described by the analytic Schwarzschild solution, which can
be parametrized by their central pressure, $p_\text c$. Starting from the
static end of the Maclaurin sequence  ($A=1, p_\text c=0$), let us increase the
central pressure until it becomes infinite. This sequence corresponds to a mere
point in the lower right corner of Fig.~\ref{Vollbild} and to the segment from
(a) to (b) in the schematic diagram of Fig.~\ref{Schema}. Configurations with
infinite
central pressure are also clearly boundary configurations. We can parametrize
these by the normalized angular velocity and follow the sequence from
the static limit (where
the angular velocity is of course zero) up to its maximal value marked by a
mass-shedding limit ((b) to (c) in Fig.~\ref{Schema}). Configurations at the
mass-shedding  limit rotate with an angular velocity equal to that of a test
mass rotating at the equator on a circular orbit. They exhibit a cusp at the
equatorial rim and are thus easy to identify by the geometry of their surface.
We define a mass-shedding parameter $\beta$ as in \cite{AKM3} by
\[ \beta:= -\frac{{r_2}}{{r_1}^2}\lim_{\varrho \to r_2}\,
           \zeta_{\text s}\frac{d\zeta_{\text s}}{d\varrho}, \]
where $\zeta=\zeta_{\text s}(\varrho)$ represents the boundary of the
configuration (see Section~\ref{Equations}). The mass-shedding parameter takes
on the value $\beta=0$ in the mass-shedding limit and $\beta=1$ for Maclaurin
spheroids. Fixing the value $\beta=0$, we proceed along the mass-shedding
sequence by reducing the central pressure until we reach the Newtonian limit
$p_{\text c}=0$ ((c) to (d)).

We can trace the last part of this sequence in the enlargement
in Fig.~\ref{Vollbild},
i.e.\ follow the red curve as it comes in from the top. This curve
crosses the Maclaurin sequence in
the parameters plotted there at a value for the radius ratio of $A \approx
0.195$. The two configurations at this cross-over point are however clearly not
the same since one is a relativistic star rotating at the mass-shedding limit
whereas the other belongs to the Newtonian Maclaurin sequence. In some other
parameter diagram these two configurations would correspond to distinct points.
This and various other intersections in Fig.~\ref{Vollbild} are
merely a consequence of the fact that the configurations considered in this
paper depend on two parameters, but not always in a unique way.

We resume our journey along the boundary curves by following the
Newtonian sequence, $A_1^+$, from (d) to (e) (parametrized again by $p_\text
c=0$ for example) by increasing $\beta$ until we reach the Maclaurin
curve at the value $\beta=1$ and find ourselves precisely at the bifurcation
point $A=A_1$. This Newtonian sequence and indeed all Newtonian
sequences bifurcating from the Maclaurin sequence at the first ten $A_n$ were
studied in \cite{AKM2} in which a depiction of the shape of the
configurations along these sequences was also provided.

\begin{figure*}
 \includegraphics[width=0.9\textwidth]{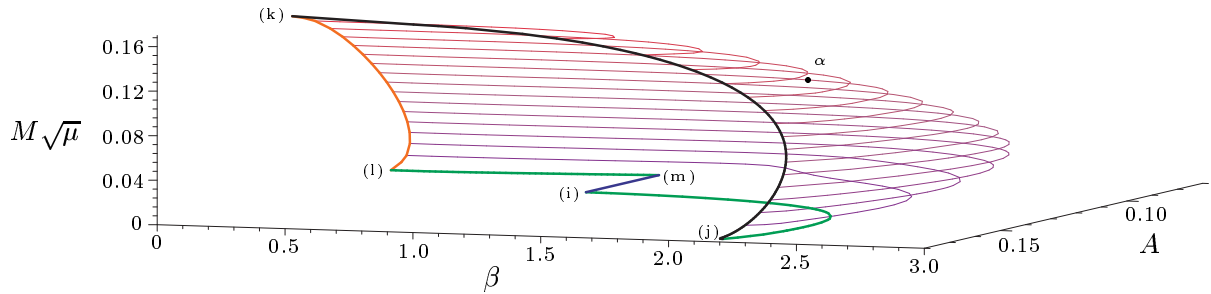}
 \caption{\label{3d} A three-dimensional parameter diagram depicting all of
class~III. Lines of constant $M\sqrt{\mu}$  beginning with the Newtonian limit
$M\sqrt{\mu}=0$ (green and blue curves) are drawn in steps of 0.01.
The colour coding of the boundary sequences corresponds to that of
Fig.~\ref{Vollbild}.}
\end{figure*}

Except for the Schwarzschild solution
itself, we consider all the boundaries in the classes to be open,
i.e.\ the limiting sequences themselves are not considered to belong to a
solution class. In the case of the Newtonian limit or the limit of
configurations with infinite central pressure, it seems appropriate to exclude
such configurations from the physically permissable solutions. In the case of
the mass-shedding limit or the two-body limit that will arise
in later classes, this choice is a mere matter of convention.
The extreme Kerr Black Hole, which will present itself in the next section, is 
also taken to be an open boundary.

Having circumscribed the generalized Schwarzschild class by
specifying its boundary configurations, it is possible to discuss
those configurations making up this class.
Amongst the most interesting properties of this first class is
that there is an upper limit on the
attainable mass, which is much higher than that known from the static solution.
The maximal gravitational mass $M\approx 0.19435/\sqrt{\mu}$ is roughly 34\% greater than for the corresponding non-rotating configuration and is reached at (c).

  \subsection{Class II: The Generalized Dyson Ring Class}\label{ClassII}
A portion of the generalized Dyson ring class was explored in \cite{AKM4}.
Here we extend that work and define the entire class by navigating its boundaries as with class~I. We begin by placing
ourselves on the Maclaurin curve at the point $A=A_1$ and proceeding
along the sequence $A_1^-$ by
decreasing the value of $A$ for $M\sqrt{\mu}=0$. It turns out that
configurations along this sequence pinch together in the centre ($A=0$),
marking the transition from a spheroidal to a toroidal topology. This behaviour
was predicted by \cite{Bardeen71}, who also correctly surmised that ring
configurations are connected to the Maclaurin sequence via the bifurcation point
$A_1$. \cite{Wong74} as well as \cite{Eriguchi81}
studied such configurations and their connection to the Maclaurin sequence,
aspects of which were clarified in \cite{AKM2}.

If we define $A$ for toroidal topologies to be the negative ratio of the inner
to outer coordinate radius of the ring, i.e.%
\[ A:=\frac{r_1}{r_2} \left\{ \begin{array}[c]{ll}
       \begin{array}{lll}
                      r_1&=&\text{polar radius}\\
                      r_2&=&\text{equatorial radius}
        \end{array} &
           \text{spheroidal}
        \\ \\
        \begin{array}{lll}
                 -r_1&=&\text{inner radius}\\
                     r_2&=&\text{outer radius}
        \end{array} &
            \text{toroidal}
        \end{array} \right. \]
then we can continue to decrease its value until we finally
arrive at $A=-1$. This limit, about which Dyson expanded his ring solution
\citep{Dyson92,Dyson93}, is
problematic however, since the ``corotating Newtonian surface potential'' tends
to minus infinity if the Newtonian mass and the radius $r_2$ are chosen to be
strictly positive and finite. The open circle at (f) in
Figs~\ref{Vollbild} and \ref{Schema} indicates the subtle and interesting jump
to the extreme Kerr Black Hole limit ($Z_0\to\infty$).

Although the horizon of the extreme Kerr Black Hole viewed from the ``outside''
world is a mere point at the origin of the coordinate system used here, there
exists an ``inner'' world (if the limit is taken in rescaled coordinates, see
\citealt{AKM4,Meinel02}) in which we can increase
the value of $A$ until we come to a mass-shedding limit ((f) to (g) in
Fig.~\ref{Schema}). Following the mass-shedding sequence by decreasing the
relative
redshift, $Z_0$, we cross back over to spheroidal topologies and arrive at the
Newtonian mass-shedding configuration for a redshift of zero ((g) to (h)).
Following the Newtonian sequence $A_2^+$ by increasing the value of the
mass-shedding
parameter, $\beta$, we arrive at the bifurcation point $A_2$ of the Maclaurin
sequence and have completely circumscribed the generalized Dyson-Ring class
((h) to (i)).

The configurations in this class, in contrast to those of the generalized
Schwarzschild class, contain no upper bound for the mass. This surprising fact
would only be of astrophysical relevance if ring configurations turned out to be
stable.%
\footnote{We intend to investigate the stability of such configurations in the
future, but preliminarily suppose that they are unstable.}

Another salient feature of this class is that a continuous transition to an
(extreme Kerr) Black Hole is possible. This can be compared to the first class
for which configurations reached the boundary of infinite central pressure for
finite values of the redshift, which has a global maximum of $Z_0\approx7.37825$
in the generalized Schwarzschild class. Note that the {\it extreme} Kerr Black Hole is the only candidate for a Black Hole limit of rotating fluid bodies in equilibrium \citep{M04}.

  \subsection{Class III: The First Generalized Core-Ring Class}

In this section, we turn our attention to the hitherto unexplored class~III and locate its boundaries, depict it in a three-dimensional parameter space and provide a representative example of one of its configurations. Having arrived at $A=A_2$
from class II by increasing the value of $A$ (cf. Fig.~\ref{Vollbild}), we
proceed
into the third class along the Newtonian sequence $A_2^-$ by placing ourselves
at
$A=A_2$ and increasing the value of this ratio further. The configurations take
on a central bulge shape, but begin to pinch in toward the outer edge. Finally
the indentation pinches together yielding a ring just barely in contact with a
central core corresponding to the path from (i) to (j) in Fig.~\ref{Schema}
(see Fig.~7 in \citealt{AKM2}). Although it would be possible to consider the
separation of core and ring, we are not interested here in studying a many-body
problem and restrict our attention to single, homogeneous bodies. Thus this
``pinching together'' marks an end-point of the Newtonian sequence. We can
follow the sequence of configurations on the verge of forming a two-body system
by allowing the mass to increase. This sequence ends in a mass-shedding limit
for the configuration with a global maximum for $M\sqrt{\mu}\approx 0.16$ in this class ((j) to (k)
in Fig.~\ref{Schema}). Placing ourselves on the mass-shedding sequence, we can
allow the mass to decrease until we reach the Newtonian limit ((k) to (l)).
From here, the Newtonian sequence can be followed to the point $A=A_3$ by
increasing the parameter $\beta$ ((l) to (m)).

As was discussed in Sec.~\ref{ClassI}, two parameters do not necessarily suffice
to describe a configuration uniquely. By adding a third
dimension, one can resolve such ambiguities. In Fig.~\ref{3d}, a plot of
the $A$, $\beta$ and $M\sqrt{\mu}$ values for the configurations of Class~III can be
found. In this plot, a given point on the two-dimensional surface in the
three-dimensional parameter space refers only to a single configuration. 
A projection of the boundary curves onto the $A$-$\beta$ plane would yield a 
picture of similar complexity to the depiction of class~III in 
Fig.~\ref{Vollbild}.

\begin{figure}
 \centerline{\includegraphics[width=\columnwidth]{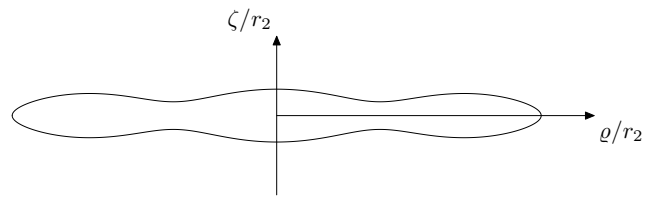}}
 \caption{\label{Sternbild} The meridional cross-section of the boundary of the
        Class~III configuration described in Table~\ref{Stern}.}
\end{figure}

An example of various physical parameters for a configuration
with the prescribed values $A=0.1$, $M\sqrt{\mu}=0.1$ can be found in
Table~\ref{Stern}. The shape of the
boundary of this configuration in meridional cross-section is shown in
Fig.~\ref{Sternbild}.

\begin{table}
  \caption{\label{Stern} Values for dimensionless physical quantities for an
    exemplary configuration in Class~III with increasing order, $m$ (see
    \citealt{AKM3}), of the numerical approximation. The quantities not yet
    referred to in this paper, $M_0$, $J$ and $R_{\text{circ}}$ are baryonic
    mass, angular momentum and circumferential radius respectively. The
    position of this configuration in
    Figs \ref{Vollbild} and \ref{3d} is marked by the letter $\alpha$.}
 \begin{center}
 \begin{tabular}{l@{\hspace{1em}}lll} \toprule
                   &      $m=20$ & $m=30$ & $m=40$ \\ \midrule
   $A$                  & 0.1 & 0.1 & 0.1\\
   $M\sqrt{\mu}$        & 0.1 & 0.1 & 0.1\\
   $M_0\sqrt{\mu}$      & 0.1093062 & 0.10930610 & 0.109306099\\
   $J \mu$               & 0.01483899 & 0.014838966 & 0.0148389661\\
   $\Omega/\sqrt{\mu}$  & 0.8627113 & 0.86271067 & 0.862710640\\
   $R_{\text{circ}}\sqrt{\mu}$ & 0.5843826 & 0.58437986 & 0.584379882 \\
   $Z_0$                & 0.4921657 & 0.49216554 & 0.492165525\\
   \bottomrule
  \end{tabular}
 \end{center}
\end{table}

  \subsection{The Further Classes}\label{Overview}
As one moves to higher and higher classes, the configurations tend to grow more
disc-like. The distance from the Maclaurin curve in the parameters of
Fig.~\ref{Vollbild} grows small although the shapes of the configurations
deviate significantly from a spheroid. We term the odd classes, beginning with
the third, ``core-ring'' classes because the configurations on the verge of forming
a two-body system contain a central body with a surrounding ring. The central
body will itself grow flat in shape and corrugated in appearance as one moves
to higher classes. Beginning with the fourth class, the even ones, on the other
hand, tend to grow concave at the centre as they ``emerge'' from the
bifurcation points. They then pinch together at the centre and proceed into the
toroidal region in which $A$ is negative. Here too, the ring configurations
grow flat and corrugated as one moves to higher classes. The corrugations
become more pronounced until finally one of them (presumably the outermost)
pinches together, whence the name ``two-ring'' classes for such configurations. The
number of corrugations of the Newtonian sequences is described in \cite{AKM2}.
There it was shown at least up to $A_{10}$ that the two Newtonian sequences
bifurcating from each $A_n$ end in a mass-shedding limit and a two-body
configuration respectively. We assume that this holds for all the bifurcation
points. Emanating from the two bifurcation points associated with a given class
are two Newtonian sequences, the one ending in a mass-shedding limit and the
other in a two-body system. Insofar as the Black Hole limit and that of
infinite central pressure are precluded from the higher classes (as indicated
by our numerical investigations), the two end-points of the Newtonian sequences
are connected by a mass-shedding and a two-body sequence. The point at which
these two configurations meet most likely marks the configuration with the
greatest mass for that class.

In order to arrive at this picture for the higher classes, we made use of the first ten Newtonian bifurcation sequences presented in \cite{AKM2} and the post-Newtonian behaviour in order to make a conjecture, which could then be verified by studying two new classes (III and IV). The properties of the remaining classes could then be extrapolated and, as such, cannot
be considered absolutely certain, but rest on firm ground.

\begin{figure*}
  \includegraphics[width=0.9\textheight,angle=90]{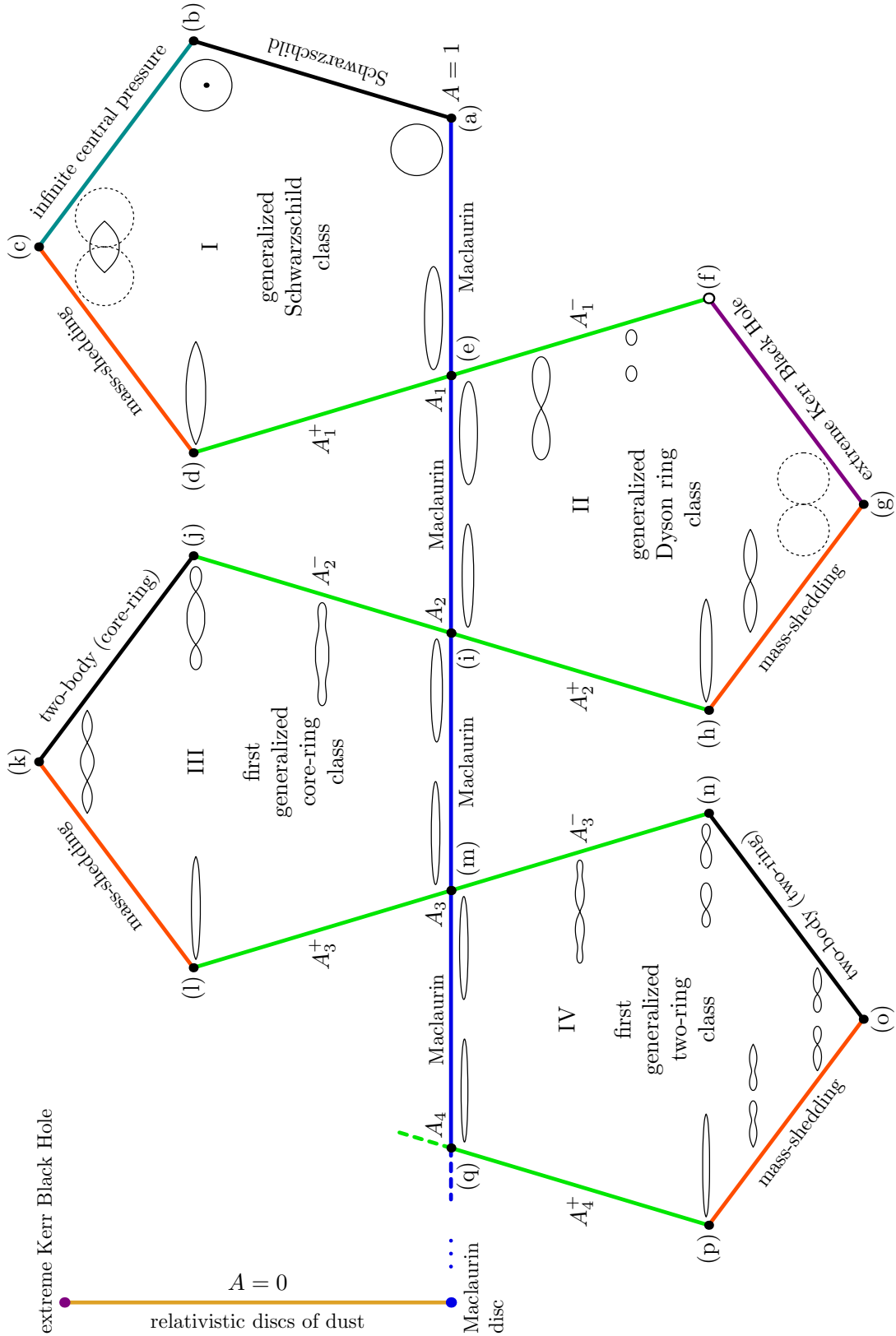}
 \caption{A schematic portrayal of the classes and their
 boundaries. A depiction of the surface of configurations in meridional
 cross-section is provided for configurations at the intersection point of two
 boundary sequences (except for the point (f), see Sec.~\ref{ClassII}). In
 addition, various surfaces are shown to help illustrate
 the transition from one corner to the next. Whenever ergo-regions arise,
 they are depicted by dashed lines (hence the dot in the centre of the
 configuration located at (b)).
 }\label{Schema}
\end{figure*}

In Fig.~\ref{Schema}, the classes are depicted schematically. As in this
depiction, two neighbouring classes have only the bifurcation point along the
Maclaurin sequence in common. The first two classes are atypical in that the
first contains the static boundary sequence and one with infinite central
pressure and the second contains a sequence of infinite redshift (extreme Kerr
Black Hole sequence). Furthermore, these are the only two classes in which we found ergo-regions. Our investigations show that they are always toroidal in topology and ``begin'' to appear within the fluid configuration. A more detailed discussion of the ergo-regions can be found in \cite{AKM3,AKM4}.

The definition of the classes given in Section~\ref{Maclaurin} relies on the
fact
that neighbouring classes have only the bifurcation point along the Maclaurin
sequence $A_n$ as a common boundary. As was already mentioned, the
post-Newtonian approximation strongly suggests that any sequence of constant,
strictly positive mass, parametrized by $\beta$ and $A$ for example, will be
deflected away from the singular point $\rdr=A_n$ and ``back into'' the
original class. This is exactly what is observed numerically.
\begin{figure}
 \includegraphics[width=\columnwidth]{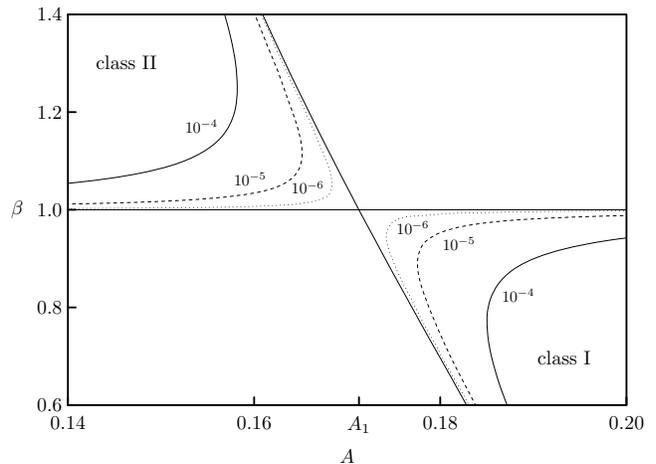}
 \caption{\label{Epsbild} The relationship of $\beta$ to $A$ for sequences
	    of constant mass in the vicinity of the bifurcation point associated
	    with $A_1$ (the point (e), cf.\ Fig.~\ref{Schema}). The solid line
	    represents configurations with $M\sqrt{\mu}=10^{-4}$, the dashed
	    line with $M\sqrt{\mu}=10^{-5}$ and the dotted line with
	    $M\sqrt{\mu}=10^{-6}$. The thin horizontal line represents the
	    Maclaurin sequence and the thin line crossing it is made up of the
            Newtonian
	    bifurcation sequences $A_1^+$ and $A_1^-$ (see Fig.~\ref{Schema}).
	     Note the gap
	    between the sequences of class~I and class~II.}
\end{figure}
In Fig.~\ref{Epsbild},  $\beta$ versus $A$ values are plotted for sequences of
constant mass. Numerical investigations show that sequences belonging to
different classes are completely disjoint although the gap in the vicinity of a
given
$A_n$ grows small for decreasing mass and approaches the point $(A_n,1)$. This
feature, which was already conjectured by \cite{Bardeen71},
can be seen distinctly in the figure around $A_1$. Such a gap was
observed numerically around $A_2$ and $A_3$ as well. A comparison with 
Fig.~\ref{3d} shows why the curve $M\sqrt{\mu}=0.01$ of that plot deviates so
markedly from the Newtonian boundary curves. If one were to plot such contour lines for
increasingly diminishing values of $M\sqrt{\mu}$, the tendency to approach the Newtonian 
limit would be evident.

 \section{Discussion}
Numerical evidence suggests that as one proceeds to flatter configurations
(higher classes), the departure from the Newtonian sequence grows small. This
is reflected in the fact that boundaries associated with highly relativistic
attributes such as infinite central pressure or the transition to a Black Hole
are presumably absent from all classes upwards of class~II and is a result of
excluding many-body configurations. If one were to abandon this
restriction and choose some (arbitrary) relationship for the rotation of the
various segments, then it would again be possible to reach ``relativistic
boundaries'' most likely. With this restriction in place, however, the only
path to the disc of dust passes through an infinite number of bifurcation
points and sees the configurations growing ever flatter and more and more
corrugated until one lands necessarily at the Newtonian Maclaurin disc. Only
once this limit has been reached can one proceed again into the
relativistic regime, i.e.\ the relativistic disc of dust \citep{NM95} and indeed
reach the extreme Kerr Black Hole. This picture
reinforces Bardeen's speculations \citep{Bardeen71}, who wrote that the
singularities in the post-Newtonian expansion at the bifurcation points ``may
forbid the existence of any highly relativistic, highly flattened, uniformly
rotating configurations which are simply connected.'' He went on to conclude
that the ``only acceptable model of an infinitesimally thin, uniformly rotating
disk in general relativity may be one that is made up of an infinite number of
disjoint rings.''

The complicated and highly non-linear set of equations describing axially
symmetric, stationary, uniformly rotating fluids of constant
density can only be solved approximately or numerically. As such, it is
difficult to find strict results describing the solution set. It is conceivable,
for example, that there exist solutions to these equations that possess no
Newtonian limit (and must thus be disconnected from the classes of solutions
introduced here). It is also conceivable
that Newtonian solutions exist that have not yet been found, but serve as limits
to relativistic solutions.

Although our numerical considerations do not preclude the possibility of the
existence of undiscovered solutions, we presume to conjecture that the classes
presented here cover the entire solution space. Attempts to prove or disprove
this conjecture are bound to lead to innovative insights into the nature of
Einstein's equations and novel methods for probing their structure. The modification to the picture drawn in this paper resulting from a change in the equation of state will be discussed elsewhere.

 \section*{Acknowledgments}
   This research was funded in part by the Deutsche Forschungsgemeinschaft
   (SFB/TR7--B1).

 \bibliographystyle{mn2e}
 \bibliography{Reflink} 

\begin{thebibliography}{}

\bibitem[\protect\citeauthoryear{Ansorg, Kleinw{\"a}chter \& Meinel}{Ansorg
  et~al.}{2002}]{AKM1}
Ansorg M.,  Kleinw{\"a}chter A.,    Meinel R.,  2002, Astron.\ Astrophys., 381,
  L49

\bibitem[\protect\citeauthoryear{Ansorg, Kleinw{\"a}chter \& Meinel}{Ansorg
  et~al.}{2003a}]{AKM3}
Ansorg M.,  Kleinw{\"a}chter A.,    Meinel R.,  2003a, Astron.\ Astrophys.,
  405, 711

\bibitem[\protect\citeauthoryear{Ansorg, Kleinw{\"a}chter \& Meinel}{Ansorg
  et~al.}{2003b}]{AKM4}
Ansorg M.,  Kleinw{\"a}chter A.,    Meinel R.,  2003b, Astrophys.\ J.\ Lett.,
  582, L87

\bibitem[\protect\citeauthoryear{Ansorg, Kleinw{\"a}chter \& Meinel}{Ansorg
  et~al.}{2003c}]{AKM2}
Ansorg M.,  Kleinw{\"a}chter A.,    Meinel R.,  2003c, Mon.\ Not.\ R.\ Astron.\
  Soc., 339, 515

\bibitem[\protect\citeauthoryear{Bardeen}{Bardeen}{1971}]{Bardeen71}
Bardeen J.,  1971, Astrophys.\ J., 167, 425

\bibitem[\protect\citeauthoryear{Bardeen \& Wagoner}{Bardeen \&
  Wagoner}{1971}]{BW71}
Bardeen J.,  Wagoner R.,  1971, Astrophys.\ J., 167, 359

\bibitem[\protect\citeauthoryear{Butterworth \& Ipser}{Butterworth \&
  Ipser}{1976}]{BI76}
Butterworth E.,  Ipser J.,  1976, Astrophys.\ J., 204, 200

\bibitem[\protect\citeauthoryear{Chandrasekhar}{Chandrasekhar}{1967}]{Chand67}
Chandrasekhar S.,  1967, Astrophys.\ J., 147, 334

\bibitem[\protect\citeauthoryear{Dyson}{Dyson}{1892}]{Dyson92}
Dyson F.,  1892, Philos.\ Trans.\ R.\ Soc.\ London, Ser. A, 184, 43

\bibitem[\protect\citeauthoryear{Dyson}{Dyson}{1893}]{Dyson93}
Dyson F.,  1893, Philos.\ Trans.\ R.\ Soc.\ London, Ser. A, 184, 1041

\bibitem[\protect\citeauthoryear{Eriguchi \& Sugimoto}{Eriguchi \&
  Sugimoto}{1981}]{Eriguchi81}
Eriguchi Y.,  Sugimoto D.,  1981, Prog.\ Theor.\ Phys., 65, 1870

\bibitem[\protect\citeauthoryear{Gondek-Rosi\'nska \&
  Gourgoulhon}{Gondek-Rosi\'nska \& Gourgoulhon}{2002}]{GonGour02}
Gondek-Rosi\'nska D.,  Gourgoulhon E.,  2002, Phys.\ Rev.\ D, 66, 044021

\bibitem[\protect\citeauthoryear{Hachisu \& Eriguchi}{Hachisu \&
  Eriguchi}{1984}]{HuE}
Hachisu I.,  Eriguchi Y.,  1984, Publ. Astron. Soc. Japan, 36, 497

\bibitem[\protect\citeauthoryear{Lichtenstein}{Lichtenstein}{1933}]{Lichtenste%
in}
Lichtenstein L.,  1933, Gleich\-ge\-wichts\-fi\-guren Rotierender
  Fl{\"u}s\-sig\-keiten.
Springer, Berlin

\bibitem[\protect\citeauthoryear{Maclaurin}{Maclaurin}{1801}]{Maclaurin}
Maclaurin C.,  1801, A Treatise on Fluxions. In two volumes, second edn.
William Baynes \& William Davis, London

\bibitem[\protect\citeauthoryear{Meinel}{Meinel}{2002}]{Meinel02}
Meinel R.,  2002, Ann.\ Phys.\ (Leipzig), 11, 509

\bibitem[\protect\citeauthoryear{Meinel}{Meinel}{2004}]{M04}
Meinel R.,  2004, Ann.\ Phys.\ (Leipzig), 13, 600

\bibitem[\protect\citeauthoryear{Neugebauer \& Meinel}{Neugebauer \&
  Meinel}{1995}]{NM95}
Neugebauer G.,  Meinel R.,  1995, Phys.\ Rev.\ Lett., 75, 3046

\bibitem[\protect\citeauthoryear{Neugebauer \& Meinel}{Neugebauer \&
  Meinel}{2003}]{NM03}
Neugebauer G.,  Meinel R.,  2003, J.\ Math.\ Phys., 44, 3407

\bibitem[\protect\citeauthoryear{Petroff}{Petroff}{2003}]{P03}
Petroff D.,  2003, Phys.\ Rev.\ D, 68, 104029

\bibitem[\protect\citeauthoryear{Petroff \& Meinel}{Petroff \&
  Meinel}{2001}]{PM}
Petroff D.,  Meinel R.,  2001, Phys.\ Rev.\ D, 63, 064012

\bibitem[\protect\citeauthoryear{Poincar{\'e}}{Poincar{\'e}}{1885}]{Poincare}
Poincar{\'e} H.,  1885, Acta mathematica, 7, 259

\bibitem[\protect\citeauthoryear{Sch{\"o}bel \& Ansorg}{Sch{\"o}bel \&
  Ansorg}{2003}]{SA}
Sch{\"o}bel K.,  Ansorg M.,  2003, Astron.\ Astrophys., 405, 405

\bibitem[\protect\citeauthoryear{Schwarzschild}{Schwarzschild}{1916}]{Schwarzs%
child}
Schwarzschild K.,  1916, Sit\-zungs\-be\-richte der K{\"o}\-nig\-lich
  Preus\-sischen Aka\-de\-mie der Wis\-sen\-schaften, 1, 424

\bibitem[\protect\citeauthoryear{Stergioulas}{Stergioulas}{2003}]{Stergioulas}
Stergioulas N.,  2003, Rotating Stars in Relativity, Living Rev. Relativity,
  Online Publication: Referenced on June 21, 2003,
  \texttt{http://www.livingreviews.org/lrr-2003-3}

\bibitem[\protect\citeauthoryear{Wong}{Wong}{1974}]{Wong74}
Wong C.,  1974, Astrophys.\ J., 190, 675

\end{thebibliography}
 
 \label{lastpage}
 
\end{document}